\documentclass[manuscript]{acmart}

\usepackage{xspace}
\usepackage{amsfonts}       
\usepackage{amsmath}
\usepackage{nicefrac}       
\usepackage{microtype}
\usepackage{url}
\usepackage{lipsum}  
\usepackage{verbatim}
\usepackage{soul}

\AtBeginDocument{%
  \providecommand\BibTeX{{%
    \normalfont B\kern-0.5em{\scshape i\kern-0.25em b}\kern-0.8em\TeX}}}

\setcopyright{acmcopyright}
\copyrightyear{2018}
\acmYear{2018}
\acmDOI{XXXXXXX.XXXXXXX}

\acmConference[Conference acronym 'XX]{Make sure to enter the correct
  conference title from your rights confirmation emai}{June 03--05,
  2018}{Woodstock, NY}
\acmBooktitle{Woodstock '18: ACM Symposium on Neural Gaze Detection,
 June 03--05, 2018, Woodstock, NY} 
\acmPrice{15.00}
\acmISBN{978-1-4503-XXXX-X/18/06}

\AtBeginDocument{\colorlet{defaultcolor}{.}}

\newcommand{\rml}{Responsible ML\xspace}
\newcommand{\rmls}{RML\xspace}

\newif{\ifhidecomments}
\hidecommentstrue
\ifhidecomments
    \newcommand{\rev}[1]{\textcolor{defaultcolor}{#1}}
\else
    \newcommand{\rev}[1]{\textcolor{purple}{#1}}
\fi

\setstcolor{purple}

\begin{document}

\title[Non-Ideal RML]{Towards a Non-Ideal Methodological Framework for Responsible ML}

\author{Ramaravind Kommiya Mothilal}
\email{ram.mothilal@mail.utoronto.ca}
\affiliation{%
  \institution{University of Toronto}
  \country{Canada}
}

\author{Shion Guha}
\email{shion.guha@utoronto.ca}
\affiliation{%
  \institution{University of Toronto}
  \country{Canada}
}

\author{Syed Ishtiaque Ahmed}
\email{ishtiaque@cs.toronto.edu}
\affiliation{%
  \institution{University of Toronto}
  \country{Canada}
}


\begin{abstract}
Though ML practitioners increasingly employ various Responsible ML (RML) strategies, their methodological approach in practice is still unclear. In particular, the constraints, assumptions, and choices of practitioners with technical duties--such as developers, engineers, and data scientists---are often implicit, subtle, and under-scrutinized \rev{in HCI and related fields.} 
We interviewed 22 technically oriented ML practitioners across seven domains to understand the characteristics of their methodological approaches to RML through the lens of ideal and non-ideal theorizing of fairness.
We find that practitioners’ methodological approaches fall along a spectrum of idealization.
While they structured their approaches through ideal theorizing, such as by abstracting RML workflow from the inquiry of applicability of ML, they did not pay deliberate attention and systematically documented their non-ideal approaches, such as diagnosing imperfect conditions. 
We end our paper with a discussion of a new methodological approach, inspired by elements of non-ideal theory, to structure \rev{technical} practitioners’ RML process \rev{and facilitate collaboration with other stakeholders}.

\end{abstract}

\ccsdesc[500]{Human-centered computing~Empirical studies in HCI}

\keywords{Responsible ML, Fairness, Machine Learning, Justice, Ideal Theory, Non-Ideal Theory, ML Practitioners}


\maketitle

\section{Introduction}

Responsible Machine Learning (RML) is increasingly becoming a top priority for organizations, international bodies, state institutions, startups, NGOs, and anyone pursuing ML-informed decision-making \cite{deshpande2022responsible,lu2022responsible,benjamins2019responsible,fjeld2020principled,wang2023designing}.
RML is loosely defined such that any step geared towards better ethical design, development, and deployment is formulated as a ``responsible'' use of ML and allows its association with diverse values such as fairness, transparency, accessibility, privacy, and inclusion \cite{jakesch2022different,dignum2017responsible}.
Further, \rev{prior research in HCI and related fields often discuss these steps} at several levels: from inter-organizational \cite{schaich2021four,jobin2019global}, intra-organizational \cite{deng2023understanding,stahl2022organisational,ibanez2022operationalising,rakovaWhereResponsibleAI2021}, team \rev{\cite{collabprac,hartikainen2023towards}}, and individual practitioner levels \cite{rakova2021responsible,yildirim2023investigating,heger2022understanding}.
Due to this multidimensional pursuit of RML, developing and using ML responsibly is increasingly becoming a multi-stakeholder vocation from a model-centric task.

Nonetheless, practitioners with technical duties
---such as developers, engineers, data scientists, and researchers---play the central role in data processing, model building, model evaluation, and system monitoring, with support of various kinds from data collectors, data annotators, managers, leaders, domain experts, and any other stakeholder involved in the ML pipeline \cite{ashmore2021assuring}.
While a significant effort has been directed towards algorithmic approaches of technical practitioners, \rev{recent works within HCI} and related fields are increasingly focusing on how RML is or could be approached at a team or institutional level \rev{\cite{krijger2023ai,subramonyam2022solving,passiTrustDataScience2018}}, such as through the integration of ethical charters, legal tools, and technical documentation \cite{pistilli2023stronger}. 
\rev{Besides a few related works on a subset of tasks performed by technical practitioners (such as coding-related) \cite{kery2018story,kery2019towards,wang2019data}, their overall} \textit{methodological} approaches in their routine tasks---such as handling data and models, monitoring, and engaging with stakeholders---are unclear. 
Recent works have discussed how practitioners use and are supported by different ethical or fairness toolkits \cite{madaio2020co,madaio2022assessing,yildirim2023investigating}. Still, minute methodological details such as the ethical implications of their statistical choices, assumptions they believe are not significant to share with other stakeholders, implicit steps they take to operationalize organizational-level RML frameworks, etc., are under-studied.
Further, while most prior works in this area typically conduct user studies on technical practitioners along with other stakeholders such as managers, domain experts, and \rev{UX/UI researchers \cite{wang2023designing,liao2023designerly,yildirim2023investigating}}, in this study, we specifically focus on the methodological approaches of those with technical duties since technical choices and assumptions are often implicit, subtle, and how they affect other components in an ML lifecycle is under-scrutinized in prior works \cite{mitchell2021algorithmic,bao2021s,mitchell2018prediction}.

It is essential to understand \rev{technical} practitioners' methodological approaches as they inform the design of intra-/inter-organizational policies with practical challenges, considerations, and opportunities on the ground for developing ML systems responsibly throughout the entire ML lifecycle.
We interviewed 22 ML practitioners working across seven domains to understand their approaches in practice, through a lens of ideal and non-ideal methodological approaches to justice or fairness\footnote{We use justice and fairness interchangeably in this work though they are understood distinctly in some traditions \cite{velasquez1990justice,sashkin1990does}.} in political philosophy \cite{valentini2012ideal,rawls2020theory,simmons2010ideal,hanel2022non}.
Ideal theory advocates for addressing injustices by simplifications, idealizations, and abstractions from real-world constraints. In contrast, non-ideal approaches focus on addressing injustices here and now through immediately accessible practical possibilities.
We use this axis of ideal and non-ideal theories to interrogate ML practitioners' methodological approaches in practice.

We find that their approach to RML falls along a spectrum of idealization.
While they brought a structure to their approaches primarily through ideal theorizing, several of their methodological choices also echoed non-ideal approaches to fairness.
However, their latter approaches, such as diagnosing imperfect conditions and mapping high-level abstract values with technical implementations, were not systematically documented and paid deliberate attention to.
Taking a non-ideal lens, we also discuss how various motivational, institutional, and resource constraints influence these methodological characteristics.  
Finally, to address these gaps, we present a new methodological framework in our discussion designed to (a) systematically document various non-ideal approaches and (b) bring structure to the value alignment process \rev{and facilitate collaboration with other stakeholders.}
Our methodological framework is a work in non-ideal theory because it provides concrete procedures to responsibly use ML 
by taking a stance sufficiently sensitive to real-world complexities that practitioners face.

\section{Related Literature}

\subsection{\rev{Responsible ML in Practice}}
\label{sec:rel_2.1}
As \citet{varanasi2023currently} put it, RML has emerged as an umbrella term to refer to movements and discourses from a wide variety of disciplines \rev{to use AI/ML, usually in high-stakes domains, to create a positive impact.} 
\rev{HCI research around the responsible use of data and AI/ML has broadly focused on two related lines of inquiries: on understanding the practices, challenges, and aspirations of AI/ML practitioners, and on improving their support systems through guidelines, methods, toolkits, etc.  }

\smallskip
\noindent \rev{\textbf{Understanding What Practitioners Do.}}
\rev{In recent years, there has been an increasing effort within the fields of HCI, Fairness, Accountability, Transparency, and Ethics/Explainability (FATE), and Science \& Technology Studies (STS) to understand what AI/ML practitioners in these domains \textit{do} across the ML lifecycle, from data collection and problem formulation to deployment. 
Regarding data, prior research has discussed how practitioners often follow complex processes drawing on domain knowledge and subjective experiences to perform their day-to-day data-related activities such as capturing, designing, cleaning, integrating, and deciding ground truth data \cite{muller2019data,bates2016data,neff2017critique,passi2017data,pine2015politics,pink2018broken,de2016social,hohman2020understanding,muller2021designing,feinberg2017design}.
Some prior works have situated practitioners' data activities within broader organizational contexts and have argued how collaborative practices and institutional barriers and norms shape practitioners' actions \cite{khovanskaya2020bottom,wang2022whose,kapania2023hunt}.
Though many of these data-related decisions influence different stages of ML pipeline and cause downstream biases, such as unfairness due to the choice of target variable \cite{passiProblemFormulationFairness2019,obermeyer2019dissecting}, prior works have highlighted how practitioners tend to ``forget'' the complexities and uncertainties in data work and do not follow a systematic and sustainable approach to data work \cite{sambasivan2021everyone,muller2022forgetting}.}

\rev{Besides data-related activities, extensive prior research has studied other practices of data scientists and people in technically-oriented roles, such as how they write code to build and evaluate models \cite{kery2019towards,kross2019practitioners}, use tools to automate various stages of ML pipeline \cite{wang2021much,wang2019human,xin2021whither,wang2021autods}, collaborate with other stakeholders in their organization \cite{muller2019data,zhang2020data}, incorporate human-centered values \cite{muller2019human,muller2020interrogating,kogan2020mapping}, among others.
The fields of HCI and ML have also produced various techniques to make ML interactive and accessible \cite{feurer2015efficient,cheng2015flock,chang2017revolt}, and prior work in HCI have studied how people without formal training in ML such as HCI researchers, software engineers, and crowd workers use data science or ML \cite{yang2018grounding,yang2018investigating,dove2017ux,zdanowska2022study}.
However, the literature in CHI and CSCW that investigates ML practices has predominantly focused on how practitioners engage with post-hoc approaches or use external toolkits to improve/evaluate ML models and make decisions based on them \cite{hongsungsoo2020,yuan2023contextualizing}.
Recently, many works also discussed practitioners' needs to go beyond explanations to integrating domain-related evidence and disclosing contextual factors, such as the characteristics of humans behind the explanations, required for decision-making \cite{yang2023harnessing,liao2020questioning,kim2023help,bhatt2020explainable,hong2020human,liao2021human}.}

\rev{Another line of inquiry that has received significant attention is practitioners' use of fairness and ethical toolkits for RML \cite{wong2023seeing,ayling2022putting,richardsonFairnessPracticePractitionerOriented2021,leeLandscapeGapsOpen2021a,ashktorab2023fairness}.
Similar to how data is engaged, \citet{dengExploringHowMachine2022} finds that practitioners' analysis choices with fairness toolkits are heavily influenced by their personal experiences, knowledge, and beliefs and discusses how practitioners even reformulate the actual ML problem in a format supported by these toolkits.
Prior research has also focused on how practitioners employ different fairness metrics (which the toolkits implement) at different levels, such as in product, policy \cite{bakalarFairnessGroundApplying2021}, and implementation, how business imperatives shape their fairness priorities and strategies \cite{madaioAssessingFairnessAI2022}, and how they use toolkits in the context of algorithmic auditing \cite{deng2023understanding}.
In summary, while prior works primarily focused on what practitioners \textit{do} at various stages of an ML lifecycle, mostly in isolation, \textbf{we take a step back and look at their overall \textit{methodological} approach that informs these practices.}}

\smallskip
\noindent \rev{\textbf{How to Better Support Practitioners.}}
\rev{Numerous works in HCI and related fields have proposed new and improved existing tools, methods, and systems to support practitioners in various tasks. Many studies that try to understand what practitioners do at work also contribute to improving ML work practices.  
In the context of RML, we could broadly classify these support systems into three interrelated categories: toolkits, frameworks, and guidelines/suggestions for design and collaboration. 
\citet{wong2023seeing} describes toolkits as ``curated collections of tools and materials,'' and they are often designed to operationalize various high-level values such as fairness or accountability, 
help practitioners discuss difficult topics and ideas, and think more critically about the contexts and impact of their work practices \cite{elsayed2023responsible,adkins2022prescriptive,ruf2022tool}.
In addition, several prior works support, modify, and improve previously developed tools drawing on theories, empirical analysis, and the practical experiences of practitioners \cite{crisan2022interactive}. 
For instance, \citet{bhat2023aspirations} design a tool, not to replace model cards \cite{mitchellModelCardsModel2019}, but to nudge the practitioners to better comply with the model cards proposals and assess the documentation quality.}

\rev{Similar to toolkits, scholars have developed various frameworks and guidelines to approach RML at various phases of ML lifecyle for data collection and processing \cite{dunbar2021towards,taylor2015data,muller2022forgetting}, disaggregated evaluation of models, explaining model outcomes \cite{ehsan2021expanding,anik2021data,liao2023designerly}, ethical risk assessment \cite{rismani2023plane}, and accountability \cite{raji2020closing}, etc. 
Researchers also develop frameworks to evaluate ethical or fairness toolkits and guide practitioners to use them appropriately \cite{richardsonFairnessPracticePractitionerOriented2021,lee2021landscape}.
Relatedly, to inform better design and development of toolkits, \citet{wong2023seeing} discuss the gap between the imagined work of ethics and the support these toolkits provide for doing those works, and \citet{holstein2019improving} discuss the challenges and needs around fairness in general.
Another category of frameworks in this space focuses on supporting ML practitioners to effectively collaborate with stakeholders from diverse disciplines and backgrounds \cite{collabprac,hartikainen2023towards,pistilli2023stronger,subramonyam2022solving}. }

\rev{\textbf{Our paper contributes to this rich body of work at the intersection of HCI and RML by throwing light on practitioners' overall methodological approach and deriving a theoretical framework to support their RML practices.}
Further, we particularly study practitioners with technical duties 
for two reasons. First, the technical choices and assumptions made by these practitioners are often implicit and subtle, and how they affect different stages of the ML lifecycle and relate to high-level RML principles are under-scrutinized.
Second, prior research in HCI investigating the practices of different stakeholders in RML often considers them homogeneously.
Recently, there has been an increasing body of work to examine the practices of UX/UI researchers in RML (e.g., see \cite{wang2023designing,liao2023designerly,yildirim2023investigating}).
However, the minute methodological details of technical practitioners, such as the ethical implications of their statistical choices, assumptions they believe are not significant to share with other stakeholders, implicit steps they take to operationalize organizational-level RML
frameworks, etc., are under-studied.
Though prior work has examined the coding-related practices of data scientists and others in related roles \cite{kery2018story,kery2019towards,wang2019data}, these practitioners' methodological approach to RML is paid less attention to.}





\subsection{\rev{Theorizing Responsible ML Practices}}
\rev{In a study by \citet{varanasi2023currently}, RML practitioners describe their processes of handling incredibly complex challenges at the institutional and individual level as a ``hodgepodge.''
While a large body of prior empirical work uses a grounded-theoretic approach to study practitioners, some have used socio-theoretical lenses, such as value-sensitive design \cite{friedman1996value,borning2012next}, to explain parts of practitioners' methodological practices or have discussed how practitioners implicitly or explicitly employ a theoretical grounding to improve their practices (examples include \cite{varanasi2023currently,liao2019enabling,ehsan2021expanding,rismani2023plane}).
However, previous research has hardly used a theoretical grounding to engage with and support RML practitioners' overall methodological approaches.
This is particularly challenging to do for technical practitioners as they have to navigate the interplay of their organizational structures and algorithmic responsibility efforts with relatively little guidance \cite{rakova2021responsible}.}

\smallskip
\noindent \rev{\textbf{Overfocus on Fairness.}}
\rev{On the other hand, prior work has discussed that, for RML, practitioners are more likely to prioritize \textit{fairness} over other values, such as transparency, privacy, and safety, and frame responsible ML as a \textit{fairness work} \cite{jakesch2022different,laufer2022four}.
In this frame, the approach is largely computational where ``fairness'' typically overlaps with an assortment of notions such as ethics, explainability, transparency, and accountability \cite{birhane2022values,weinberg2022rethinking}.}
Altogether, \citet{fazelpour2020algorithmic} observe this fairness frame of RML largely follow \textit{ideal principles} in their theorizing about normative fairness prescriptions. 
Ideal theorizing of justice or fairness starts by invoking a number of simplifying assumptions about the world to derive a set of idealized guidelines about how a phenomenon of interest should function \cite{rawls2020theory,farrelly2007justice,simmons2010ideal}.
\rev{In the context of RML fairness frame,} \citet{fazelpour2020algorithmic} explain that prior works typically follow a three-step process: (a) define in an abstract sense what ``fairness'' or other values should mean ideally (b) express this ideal in a mathematical form, and (c) minimize the above metric at different stages of the modeling process.
This process, inspired by ideal theorizing, is mathematically appealing since it offers a target state and supports quantifying discrimination by measuring the deviation from the presumed ideal target state.

However, it is argued that theorizing about fairness through the lens of idealized principles will insulate one from the most pressing concerns of society and conceal the non-ideal considerations in the real world \cite{farrelly2007justice,anderson2010imperative}, systematically overlook injustices not within the scope of operationalized notions, provide misguided normative prescriptions, and evade offering guidelines to address complex questions in practice \cite{wiens2015political,hanel2022non}.
Further, applying fairness metrics is not the only approach practitioners employ for RML. Their methodology begins with data collection and goes till monitoring models after deployment.
\rev{Prior work has not explicitly discussed how ideal theorizing could explain practitioners' methodology or how practitioners implicitly or explicitly follow ideal theorizing in different stages of an RML workflow.}

\smallskip
\noindent \textbf{Perspectives from Non-Ideal Theories.}
Non-ideal theorizing, in contrast to ideal theories, \textit{``begins with a diagnosis of the problems and complaints of our society and investigates how to overcome these problems''} \cite{anderson2010imperative}. The non-ideal theory seeks to understand the underlying causal explanations of the problem, explores the possibilities of subsequent actions, and brings in the voices of multiple actors in our deliberations about what the demands of fairness are \cite{farrelly2007justice,goodin2003reflective,wiens2015political}.
In this perspective, our choices and actions are not conceived as steps towards reaching an ideal state, but rather, they should determine the contextual factors and aim at identifying practical strategies to overcome immediate unfairness.
Ideals or standards, in a non-ideal perspective, function as hypotheses that are tested, contested, evolved, and reflected upon to improve the conditions of the present \cite{cozzaglio2022feasibility,favara2023political,anderson2010imperative}.

Recently, \citet{fazelpour2020algorithmic} introduced the non-ideal theorizing of fairness from political philosophy to the fair ML literature and provide two general ways forward: (a) through an empirical understanding of the imperfect conditions that involve an extensive analysis of various choices, value judgments, and simplifying assumptions, and (b) through an empirically informed implementation and evaluation of interventions. 
Though promising, their guidelines are not specific that can be translated to the design and development of ML applications.
\citet{lundgard2020measuring}, in their review of justice discussion in machine learning, envisions a capability-based approach to ML fairness through non-ideal modes of theorizing, building upon the works of \citet{oosterlaken2012capability}.
Though this approach is more sophisticated, due to its inherent flexible design, it may still result in an imprecise translation of the feedback from the capability measure to the ML system. For instance, when the end user's understanding of the system misaligns with the developer's intended goals of the system, then iteratively designing and evaluating the system based on capability measures could still lead to unfairness.

Overall, it is unclear if an explicit invoking of non-ideal theories will help understand how practitioners approach RML or will provide them with a better methodological framework.
To address these gaps in theorizing and understanding practitioners' methodological approaches, we ask the following three broad research questions in our study: 

\begin{itemize}
    \item (RQ1) Under what constraints do practitioners employ RML practices?
    \item (RQ2) What are the characteristics of practitioners' methodological approaches to RML?
    \item (RQ3) What support do practitioners need to improve their RML methodology?
\end{itemize}

\section{Methods}

\begin{table}[]
\begin{tabular}{l|l}
\textbf{Characteristic}  & \textbf{Distribution}                                                                                                                                                                                             \\ \hline
Gender                   & Female: 5, Male: 17                                                                                                                                                                                               \\ \hline
Years of ML experience   & Min: 2, Median: 6, Max: 11                                                                                                                                                                                        \\ \hline
Role type                & Data Scientist: 8, Developer: 4, Engineer: 3, ML Scientist: 2, Researcher: 5                                                                                                                                      \\ \hline
Domain of ML application & \begin{tabular}[c]{@{}l@{}}Agriculture: 4, Environment: 3, Education: 4, Finance: 3, Human Resources: 3, \\ Public Health: 7, Public Service Delivery: 3\\ (some practitioners were involved in multiple projects in different domains)\end{tabular} \\ \hline
Institution type         & Academia: 5, Government Bodies: 3, For-profit industry: 10, Non-profit industry: 4                                                                                                                                                       \\ \hline
Institution scale        & Large: 9, Medium: 8, Small: 5                                                                                                                                                                                     \\ \hline
ML type                  & \begin{tabular}[c]{@{}l@{}}Reinforcement learning: 4, Supervised learning: 12, Unsupervised learning: 9\\ (some practitioners were involved in multiple projects with different objectives)\end{tabular}               \\ \hline
Location                 & Global North: 8, Global South: 14                                                                                                                                                                                 \\ \hline
\end{tabular}
\caption{\textbf{Participant Information}}
\label{tab:methods}
\end{table}
To answer our research questions, we conducted semi-structured interviews with 22 ML practitioners during June-August 2023. We conducted all our interviews in English. 

\smallskip
\noindent \textbf{Participant Recruitment.}
We recruited participants through a combination of purposive and snowball sampling \cite{sharma2017pros}. We sent out our recruitment message, asking about their technical duties at work through email and social media announcements, internal forums, newsletters, and mailing lists. 
Of the responses we received, we excluded respondents (a) whose roles and job requirements did not have technical duties such as performing data analysis or building, shipping, and monitoring ML models, (b) who had less than a year of professional experience, such as students involved in RML-related tasks for their theses or internships, and (c) who wanted to execute the responsible side of ML in their work but were not practically involved at the time of our recruiting. 
However, we considered respondents from academia who performed \textit{industry-level ML tasks} that required them to develop and deploy ML responsibly in various domains.  
We got approvals from our institution's ethics board before commencing the study. 

\smallskip
\noindent \textbf{Participant Information.}
We interviewed 22 ML practitioners for about two months after excluding respondents who did not suit our study purposes. 
Our participants came from 10 different countries across the world, where 14 were located in the Global South and the rest (n=8) in the Global North. 
Most of our participants were industry practitioners whose organizations ranged from small-scale startups to large multinationals.
Five participants were associated with academic institutions, and three worked with organizations undertaken by government bodies.
Our participants had at least two years of experience developing ML systems and a median experience of 6 years. 
We summarize our participant information in Table \ref{tab:methods}.

\smallskip
\noindent \rev{\textbf{Participant Roles.} }
\rev{A few of our participants (n=5) held senior positions, such as lead data scientists or professors, or had multiple responsibilities, such as managing a small team in addition to research.
While the nature of all the tasks they do would differ across these roles, our interviews specifically focused only on RML-related on-the-ground tasks (such as data processing, model building, and evaluation) they performed at the time of the interviews or a few years before.
As the field of RML itself is still developing, almost all our participants started to focus more on RML-related strategies only roughly in the past 5-6 years.
Similarly, though participants from academia varied from PhD students to professors, all of them were actively involved in industry-level ML tasks, sometimes in collaboration with state or private organizations.
In summary, our participants' nature of work often involved more than one or all of the four stages---data management, model learning, model verification, and model deployment---discussed in \citet{ashmore2021assuring}'s survey of practices followed in typical machine learning lifecycle.}

\smallskip
\noindent \textbf{Interview Procedure.}
We obtained informed consent from all our participants well before the interviews (and again before beginning the interviews). Also, we informed them of their rights to withdraw from the study at any time without reason.
We conducted and recorded all our interviews via audio/video calls upon consent (either with or without video), and we manually took notes of their responses for participants who wanted to be kept from being recorded.
Our interview questions were categorized into three themes, trying to elicit responses to each RQ, with some questions evoking responses for multiple RQs.
After a few rounds of interviews, we started receiving less information about the constraints regarding methodological choices, so we revised our interview script to ask questions that gave us more profound insights about practitioners' methodological characteristics and the methodological support they needed.
We stopped conducting interviews once we did not receive new insights.
Our interviews lasted for about one hour, and we compensated our participants C\$30 for their contribution.

\smallskip
\noindent \textbf{Data Analysis.}
We first transcribed all our interviews through an automated transcription software that converted audio to text locally using open-source ML models. We then went over these texts and manually corrected them for any misinterpretations. We stored all recorded interviews, notes, memos, and other study materials in the first author's institutional cloud storage, with access restricted only to co-authors. We removed any personally identifiable information about participants to preserve anonymity.
We carried out an abductive analytical approach to find themes in our data \cite{timmermans2012theory}. We used concepts from ideal and non-ideal theories as foundations to understand practitioners' methodological approaches but were also flexible to allow newer insights and perspectives to emerge inductively. 
We iteratively refined our codes over several rounds of careful reading and coding of the transcripts, guided by ideal and non-ideal theories.



\section{Findings}

Our findings are divided into three sections addressing each research question in order. We begin by discussing key constraints that situate ML practitioners to adopt various methodological choices and considerations (RQ1). We then discuss the defining characteristics of their methodological approaches to Responsible ML (RQ2). The final section details mechanisms that would support their methodological inquiry (RQ3).

\subsection{Constraints Determining Methodological Choices}
\label{sec:constraints}

In his non-ideal theorizing of justice, \citet{wiens2015political} argues for factoring in various real-world constraints--such as institutional, economic, and motivational--in determining political possibilities to overcome injustices. Similarly, in the case of \rml, we find that practitioners are constrained by several intrinsic (for instance, their mental models or affective biases) and extrinsic factors (of their environment) that shape their methodological choices for the responsible use of ML. Below, we discuss three broad categories of such constraints. We discuss these categories separately for explanatory purposes, but we note that these divisions influence each other, and boundaries are fuzzy.

\smallskip
\noindent \textbf{Institutional Constraints.} 
\rev{While prior works in HCI have discussed some of the practitioners' institutional constraints in the context of collaborative practices between or within teams \cite{varanasi2023currently,subramonyam2022solving,zhang2020data}, we discuss the constraints our participants faced through a \textit{methodological lens} using ideal and non-ideal theories.}
Restricted control over data provenance was highlighted as one of the significant determinants of \rmls methodological choices. Eight participants shared that data reaches them in secure servers for some of their projects with government organizations, which are then processed for model training. They have little control over analyzing, reviewing, or altering the data collection/generation process or have scarce knowledge about assumptions made before the data reaches them. Consequently, practitioners develop heuristics to address data quality issues such as missing, inaccurate, or biased data. While some participants (n=3) shared that they frame these heuristics based on informal field observations, the rest base decisions on personal intuitions. 
Below, P09, an ML research engineer working on public health, shared how they make several subjective choices when using pre-processing fairness methods \cite{kamiran2012data,calmon2017optimized} to alter the data-generating process:

\begin{quote}
    \textit{``As we mostly work on third-party data, we have little control over it [data]. Consider the pre-processing fairness techniques we use, such as relabelling, reweighting, etc. We assume that labels of some data points [about human demographic and clinical details] can be relabelled without actually knowing if that is okay. We mostly do it for statistical convenience.''}
\end{quote}

Even if data bias is supposedly addressed, P17, a data scientist who worked on several ML projects with state organizations across the world, noted that \textit{``the bias in the original data still exists; if a different company uses the same data, there is a good chance that they follow a different pre-processing method, make different assumptions about the data-generating process, and get different results.''} Several practitioners, especially those whose clients were government bodies, highlighted that such reproducibility crises surface when they are constrained by institutional policies on meta-data accessibility, such as how the data was collected or annotated. Participants shared that knowing a little more---such as why some features were collected/not collected or why some features were measured in a particular way, etc.---could provide them with contextual information to decide what pre-processing steps would be more appropriate. Instead, as several participants shared (n=9), the currently dominant approach is to try out many methods and choose the one that gives the least biased results and aligns with practitioners' prejudices. 

Further, prior work has observed that top-down institutional structures constrain ML practitioners' exploratory value mapping and push them to align their methods to organizational/business values \cite{chen2021beyond,madaioCoDesigningChecklistsUnderstand2020,varanasi2023currently}. In addition to these observations, in our interviews, we also find that hierarchical structures limit the communication of ML practitioners' methodological choices to other stakeholders. 
Almost all participants (n=17), except a few working in academia and engineering-driven companies, shared that their bosses were not available to consume information about methodological choices they made that could have fairness implications; their jobs were reduced, as P16 pointed out, \textit{``to checking if the data is okay.''}
We further observe that, even in \rev{relatively} flat-organizational settings such as academia or research institutions, non-technical stakeholders such as domain experts or program managers largely do not show interest in understanding the implications of technical and methodological considerations to downstream tasks.
A few participants (n=4) shared that they are sometimes unsure if their methodological choices implicitly affect model outcomes because they do not receive critical comments or feedback.
P05, a senior data scientist working on LLMs for a human resource management company, shared that

\begin{quote}
    \textit{``Communication with non-technical audiences is very important. I should not use any Deep Learning jargon and not even use terms like distributional shifts, etc. In one instance, when I tried to discuss why I processed the data like this using SHAP [an XAI tool], they were not interested in such conversations. I insisted that I assumed something about the data, which might impact how our model predicts for some input data points. However, they said all these were okay as long as the model performed well and SHAP produced good plots.''}
\end{quote}

\smallskip
\noindent \textbf{Motivational Constraints.}
One of the distinguishing features of \citet{wiens2015against}'s non-ideal account of justice is his introduction of motivational constraints in assessing what political actions are feasible to overcome injustices. 
We find this very relevant to \rml, where practitioners' motivations influence their methods. Consider the use of checklist-type toolkits that many ML practitioners use widely in their work. \rev{Prior research in HCI} has discussed both the pros and cons of checklists: while they provide non-specific guidelines, they are useful in invoking critical and reflexive thoughts to improve ML systems \rev{\cite{wong2023seeing,yildirim2023investigating,wong2023seeing}}. In other words, the usefulness of checklist-type toolkits depends on how \textit{motivated and invested} practitioners are to ask further critical questions. However, seven participants in our interviews shared that ML practitioners often have no incentive to engage in such discussions. P06, a research scholar in academia working on AI and health, shared their thoughts as follows:  

\begin{quote}
    \textit{``I think they [checklists] are good critical questions about your design. But to actually incentivize someone to engage critically in their own design is more than just asking them a specific question. I think there's some individual incentive and motivation that they have to have an interest or drive to sort of do that... I think for most people who engage in this work, there's no incentive for them to engage in these questions before they start their work. There's really no incentive afterward either''}
\end{quote}

Further, as ten participants shared, after some point, they get exhausted after pushing with less or no support. P02, a data scientist who worked with state organizations to improve several cities' road connectivity and infrastructure, was motivated to lead a team in following participatory design practices without external support for several months. However, they noted that after a point in time, their team members became tardy and unmotivated to pursue further, so end up making simplifying assumptions. Similarly, in another instance, P04, a data scientist working for an ML consultancy company, shared that \textit{``unless fairness comes as a requirement, we don't do it. We mostly stop at unit testing or if there is any bad press or legal issues.''}


\smallskip
\noindent \textbf{Resource Constraints.}
While the cost of training and hosting ML models are getting increased attention with a rise in the use of LLMs \cite{strubell2019energy,bender2021dangers,liang2022holistic}, the type and proportion of these costs directed to ``de-bias'' the data or retrain/re-evaluate the models is not given due attention. 
As discussed in our literature review, the fairness approaches of participants in our study were predominantly computational in nature. Practitioners, especially those from academia, NGOs, and research institutions, shared that ensuring the fairness of ML systems (computationally) is a time-consuming, resource-intensive, and uncertain process.
While most participants (n=16) agreed that ensuring fairness requires good computing infrastructural support as the models need to be re-trained and re-evaluated with different parameters, a few participants (n=5) also shared that most projects in the real world require continuous monitoring over time as a fair model once may become unfair in the future: new biases emerge, or old fairness evaluations need to be reassessed. 
P03, an ML team lead working on public health, expressed how resource constraints impacted their fairness methodology:

\begin{quote}
    \textit{If we had enough time, resources, and control, we would start fairness work from the start [hinting at data collection stages]. However, in reality, we can't perform fairness analysis forever; in the end, we have to provide some models as the government won't wait for long... getting something quick and good enough out there is important. So speed, yeah, speed [implying how quickly a model is deployed] is an important thing, even more than responsibility [implying RML practices]. That sounds a little weird, but that's how it is.}
\end{quote}


\subsection{Defining Characteristics of RML Methodology}
\label{sec:characteristics}

\citet{fazelpour2020algorithmic} observe that the common framing of computational approach to fairness aligns with the ideal modes of theorizing justice or fairness \cite{simmons2010ideal,stemplowska2012ideal}, which typically follows three stages: operationalizing an ideal conception of the world, developing quantitative metrics to capture the deviation of the observed world from this ideal standard, and enforcing interventions to minimize the measured deviation.
However, in our interviews, we find that practitioners do not follow such a definitively straightforward process, but rather, their methodological approaches fall at different points in a spectrum of idealization, where strict ideal and non-ideal theory fall at two opposite ends.

\subsubsection{Structuring through Ideal Theorizing}
\label{sec:ideal}
Our participants did not follow any uniform structured methodology \rev{as observed in a few prior works within HCI \cite{varanasi2023currently,rakovaWhereResponsibleAI2021}}. However, the process of structuring their approaches often aligned with ideal modes of theorizing fairness.  
Further, this happened at two levels. First, practitioners abstracted RML workflow from the inquiry of the applicability of ML to a specific problem. 
Second, practitioners interpreted, conceptualized, and operationalized principles and values in their RML workflow through idealizations and abstractions. 

\smallskip
\noindent \textbf{Abstracting ML Applicability.}
We define determining the \textit{applicability of ML} to a specific problem, based on empirical results, as a critical inquiry of the methods employed and assumptions made to realize the intended goals of the ML system.
This inquiry is intertwined with the interrogation of fairness or ethical considerations, as different methods and assumptions are likely to have some form of fairness implications. 
Prior work has shown that micro-decisions made in routine steps followed in ML applications, such as construct operationalization, missing values imputation, feature transformation, and setting evaluation standards, are related to the fairness of ML systems \cite{saxenaRethinkingRiskAlgorithmic2023,fernando2021missing,jarrahi2022principles}.
We call this the applicability inquiry because such an analysis gives a qualitative indication of the extent to which ML is applicable to a problem.
When too many assumptions have to be made to realize a particular goal, for instance, if a dataset with a large number of high-cardinal categorical features has to be processed to an extent where the features' contribution to model outcomes do not make sense, then the applicability of ML is in question. 

We observed that practitioners' first level of methodological abstraction is to implicitly or explicitly dissociate the applicability inquiry from ethical or fairness inquiry. 
By ideal theorizing RML approaches, most participants (n=16) focused on the intervention of fairness metrics but often did not question the object of intervention: the application of ML to a problem. 
Only three participants reported that they carefully considered the ethical implications of each step they took in modeling.
As a result, practitioners typically did not account for details of the assumptions and choices at each routine steps and their relations to fairness.
Echoing non-ideal theorists' critique of ideal theory-based methods, we find the above abstraction too ``idealistic'' and not reflecting the empirical reality under which the model was built.

Several participants working in industrial applications of ML (n=7) shared that if they can deliver models that are somewhat better than the baselines, they will be fine. As P05 mentioned, \textit{``so in business, you always think in terms of good, bad, ugly, like it's about  reducing your area of concern in such a way that you are able to deliver value, even if it's 30\% or 40\%, it's good if it's still better than the current process.''}
However, if the inquiry of the applicability of ML to a problem is pursued, several questions about methodological assumptions and their alignment with intended goals arise.
For instance, even if a model improves the lives of one sub-group of people but does not make any change to another, as P06 noted, \textit{``it still creates two tiers in the system,''} and may require further deliberation.
However, practitioners abstract out such queries and do not juxtapose them with fairness considerations; they apply ML by making assumptions, get the system out, and then pursue fairness inquiries later.

In contextualizing why this happens, we note that various institutional, motivational, and resource constraints we discussed in section \ref{sec:constraints} influence practitioners in abstracting the inquiry of applicability of ML.
In the below excerpt, note how P14, a lead data scientist working on building ML solutions in low-resource environments, implicitly abstracted the ML applicability inquiry from ethical/fairness inquiry due to resource constraints:
\begin{quote}
    \textit{``Smartphones are less pervasive [in their context], so our model has to be shrunk to feature phones. Model accuracy gets affected in this process, and fairness there becomes less of a priority, or we are unclear how to do it. The goal is to create a decent model first and then focus on RAI issues.''}
\end{quote}

\smallskip
\noindent \textbf{Ideally Conceptualizing RML Process.}
As practitioners abstracted the ML applicability inquiry, they also referred to some form of guidelines to support their fairness or ethical inquiry. 
For some participants (n=7) in mature institutional settings, these guidelines were very structured and formal, for instance, in the form of RAI workbooks or legal charters \rev{\cite{hind2020experiences,yildirim2023investigating}}.   
In such settings, the high-level guiding values and principles often came from the top, and practitioners played a small, unstructured contributing role.
There was no standard pattern for the rest: While some practitioners (n=4) developed guiding documents bottom-up, others had a fluid, inchoate process.
Nonetheless, guidelines documents serve as communication channels with different organizational stakeholders. \rev{As prior HCI research has highlighted \cite{wong2023seeing,holstein2019improving}, our participants often found these documents} abstract, high-level, and eliciting open-ended discussions.

Practitioners then implicitly interpreted and conceptualized the values and principles in their own ways, theorized in what ways values were related, and how values could be evaluated.
\rev{Though this process could be constrained by institutional ethical committees for those from or in partnership with academia, our participants shared that most of these ethics approvals were not difficult to obtain and the responsibility of engaging more on values formulation and operationalization were often in their hands}\footnote{\rev{we recognize that this could also indicate our sample is biased.}}. Nonetheless, this process follows the conceptual aspect of the process of ideal theory \cite{erman2022ideal}, where through conceptual work, ``we may understand more precisely which considerations are at stake when we try to identify one policy option as more just than another under real-life circumstance \cite{swift2008value}.''
However, what distinguishes practitioners' conceptualization process from the general aspect of ideal theory is that they often had their own mental constructs of this process and performed it implicitly. 
While some aspects of conceptualization were recorded explicitly, in model cards, or in other documentations, how different values and principles were \textit{mapped} to the intended goals of the ML system, and the process of mapping primarily resided in the minds of practitioners.
For instance, according to P05, practitioners have their own mental representations of what different values mean, which are often ignored provided they are aligned with the business goals:

\begin{quote}
    \textit{``Say if inclusivity is a business value [in the guiding documents] and I decide to include, say, Spanish or Hindi or some other language, but no matter how you cut it, there will always be a bottom half of languages purely based on the number of people who speak, right? So I believe they [the decisions] are still not valid from an inclusivity point of view. Sometimes, they will be correct. Sometimes they will be wrong. Who knows, right? Whereas in a business context, I think they aligned with the business goals''}
\end{quote}

Note that as P05 conceptualized a value, they also investigated different interpretations of inclusivity as a value. Also, many practitioners (n=12) explicitly shared that they examine the relative importance of multiple values and principles before operationalizing one or a few of them. This is referred to as the axiological aspect of ideal theory, where the ``value of values'', the relative importance and relationship of values, are systematically and coherently analyzed.
However, in practice, instead of serving a guiding role, this process often involved abstraction from non-ideal modeling conditions, making idealized assumptions, and resulting in misguided interventions. 
For instance, P03 shared that a compromise in fairness values for some cohorts is acceptable as long as the values of impact a good model could bring are better than the current baselines (see this work on levelling down \cite{mittelstadt2023unfairness}):

\begin{quote}
    \textit{``Even if your model is unfair, if it's doing better than baselines for all cohorts, it still might be worth deploying. So maybe your recall scores on women are worse, significantly worse, than the recall for men. But the baseline there is so bad that you should still deploy the model because you still do good for those cohorts. So, we noticed a significant reduction in variance. The nice thing is that performance for the good cohorts is still good. Like it's not as good, but it's still acceptable. And that for the weaker cohorts, it certainly increased''}
\end{quote}

While the computational metrics indeed describe an idealized and abstracted version of real-world biases, as proponents of ideal theory argue, such idealized metrics can also serve as ``counterfactual devices'' to inform actions required to address non-ideal conditions \rev{(see \cite{haslanger2012resisting,haslanger2021methods} for a discussion on how such \textit{applied} ideal theory are not always the best strategy to critique social practices)}.
For instance, pre-processing fairness metrics can be used to identify patterns of discrimination in data and inform better data collection practices.
However, most participants (n=15) did not follow such methodological practices, and their approaches aligned with what P06 mentioned as, \textit{``fairness, in this sense, just acts as a checkbox. But these are just band-aid solutions to deeper problems that are more epistemological about the actual field itself.''}

\subsubsection{Not Documenting non-ideal Approaches}
\label{sec:nonideal}
While ideal theorizing is typical when practitioners approach RAI concerns, we also find that several of their methodological choices echoed non-ideal approaches to fairness. For instance, some of them considered motivational factors of different stakeholders to decide what actions were feasible and were to be pursued (n=6), focused on concrete issues rather than trying to achieve ideals (n=9), and evaluated based on how current issues were solved and not against idealized standards (n=8), etc.
However, practitioners often performed these actions without deliberate attention and did not document most of the details associated with these practices.
Below, we discuss two predominant non-ideal methodological practices that practitioners disregarded in their documentations and record-keeping.

\smallskip
\noindent \textbf{Discounting Diagnosis of Imperfect Conditions.}
One of the distinguishing proposals of non-ideal theory is to diagnose and address the immediately accessible injustices rather than trying to achieve idealized standards \cite{valentini2012ideal,wiens2015political}. 
In some of their projects, we observed that several participants (n=12) diagnosed and attempted to address various imperfect conditions\footnote{We use ``imperfect conditions'', as also used by our participants, to denote any deviations from the expected modeling conditions. For instance, a dataset with a large number of categorical features could be an imperfect condition for some applications.} affecting different stages of an ML pipeline instead of idealizations and abstractions.
For instance, P19, an academician working on AI for agriculture, shared that they focused on values such as privacy, accountability, and trustworthiness that were specific and relevant to their application instead of referring to values such as ethics or fairness, which they said would result in ``vague realizations.''
This line of thought aligns with non-ideal theorists' arguments about how the normative priority should not be the achievement of unfeasible fairness or justice but rather that of fact-bound social ideals \cite{north2010political,galston2010realism}.
Also, note that this diagnosis of imperfections is also the first step in the inquiry of ML applicability we discussed in section 4.2.1, and it requires detailed empirical investigation and critical analysis.
Consider how P06 investigates the persistence of different troubling factors in data collection and feature engineering that could produce inaccurate, unfair outcomes in predicting stress:

\begin{quote}
    \textit{``As it stands, we have very little knowledge about how most drugs work. Say aspirin. People are still bringing up theories about how aspirin works. It's not that we lack the intellectual ability to reason out why something works. It's kind of questionable why an algorithm can, with a much smaller set of covariates and features that are available to it, be able to make a prediction that would be better. The human bodies are black boxes. We just capture heart rate, respiratory signals, sleep signals, some subjective questions about how they felt that day. But that's still like the surface of what's happening in this complex system underneath. There are so many covariates and confounders, and when trying to model that complex system, algorithms can only go so far. Maybe our understanding of that system is still limited in the past. Maybe we're not even capturing the right level of features to be important for that model.''}
\end{quote}


In the above passage, P06 highlighted some open issues from their empirical investigation of predicting stress in individuals. 
Notice how they try to address ground realities instead of building a model and showing that it behaves fairly according to fairness metrics. Addressing these immediate issues could entail addressing several fairness concerns, for instance, biases due to inaccurate operationalization of stress and related features. 
However, practitioners do not have a structured way of documenting or mapping out the imperfections they are trying to resolve, assumptions they make, and goals they are trying to achieve. 
As a result, they often discount several details of this investigative diagnosis and create simplistic pictures in the existing documentation toolkits. 



\smallskip
\noindent \textbf{Discounting Varying Functions of Toolkits.}
Our participants used a range of ethical or fairness toolkits, such as model cards, datasheets, guidelines, worksheets, explainability tools, and fairness software. They served a range of functions, including but not limited to documenting, interpreting machine outcomes, and proving accountability.
While it is a common practice to use toolkits such as explainability (XAI) techniques for different purposes, such as debugging and accountability, a few practitioners (n=4) shared that the focus is always towards the end: that is, how to use these tools to explain a machine prediction after deployment.
However, P01, an academician working on ML for education, shared that sometimes the steps taken to identify discriminatory characteristics of data or models using XAI tools need to be recorded to understand how some biases were addressed, to avoid future recurrences, and ``interpret'' XAI tools' outputs correctly.


Similarly, our participants also noted that the final face of toolkits like model cards, checklists, and guidelines-based documents that we see were often developed in stages rather than in one go. 
Seven practitioners shared that the model cards they develop or encounter at work do not give any information about how they arrived at a point and mostly have minimal informative value in practice.
For instance, P14 shared that just knowing the model architecture and risks and limitations, as featured in model cards typically, does not reveal anything about the fairness-related questions addressed; instead, what matters is how the assumptions of this architecture and data processing steps relate to the stated risks and limitations.
\rev{While a few prior studies in HCI discuss some of these issues and propose new toolkits to address them \cite{adkins2022prescriptive,hind2020experiences}, we discuss how any of these toolkit can have varying functions (that is not documented) during the ML lifecycle.}
Further, P02 and P17 noted that it is impossible to write specific ethical guidelines on the first attempt; though these documents are versioned, as P01 shared, \textit{``sometimes, developers violate a value and then satisfy it later, so knowing why and how some point in the guidelines changed is very important than just versioning.''}
P17 also mentioned that sometimes the values held by top-level stakeholders (referring to their company's management) change and some values become inappropriate, but how these values mutated was typically not recorded nor discussed.


\subsection{Supporting by Mapping Values to Interventions}
\label{sec:support}
Our participants expected a range of support, from algorithmic or statistical, such as fairness metrics for LLMs, to philosophical, such as charting the relations between diverse ethical principles.  
\rev{While the fields at the intersection of HCI and ML have produced many support systems for practitioners (section \ref{sec:rel_2.1})}, our interviews focused on \rev{less attended aspects in HCI} that are expected to improve their overall methodological approaches to RML. 
A major theme that emerged across the interviews was the need to structure the value alignment process between different stakeholders (n=18). We first discuss the issues with the value alignment process our participants faced and then discuss two directions of mapping that our participants implied.

One of the first actions that our participants working with diverse stakeholders performed was ensuring that their ML application values aligned with other stakeholders. Many participants agreed that if values are misaligned, users will not use the ML systems regardless of how good the model outcomes are.
We note that several institutional and motivational values, discussed in section \ref{sec:constraints}, impact the value alignment process here.
P07 shared that, in one of their projects, their notion of fairness and its operationalization clashed with domain experts' (from public health) notion of equity. 
Practitioners, like P07, expressed that the conception of values and what interventions are carried out to achieve those values should match with the intuitions of domain experts for a smooth application of ML.
In another instance, P01 shared how a model their team built, though producing accurate predictions of student successes/failures, infuriated a domain expert in educational psychology due to a difference in problem formulation:

\begin{quote}
    \textit{``We had a team who was presenting an algorithm they developed [to predict student success in order to provide appropriate advice]. But one of the academic advisors just ran out. She was angry. She smashed the door and walked out. And she said, like, no, I'm not doing this. This is totally ridiculous.  And it was not because the system was not well designed or because the AI was not sound. It was because it was trying to advise something that she didn't feel was appropriate advice to a student.''}
\end{quote}

\rev{Non-ideal theorists emphasize the need to theorize for a world of imperfect convergence of views and argue that people could have different views of what they are advocating for even if they are advocating for the same values \cite{schmidtz2011nonideal,sen2008idea,amartya2017we,sleat2012legitimacy}.
Similarly, participants highlighted that values were understood in different ways by different stakeholders but they }were largely unclear on how to address the disconnect between values and interventions systematically \rev{using existing toolkits or frameworks.}
As P02 shared, \textit{``it [value alignment process] mostly does not happen in a structured way; it is mostly informal. It is only coincidental that these things work.''}
While most participants resolved any value mismatch through informal, unstructured conversations, a few (n=3) acted based on feedback from survey questions, such as whether the value itself was problematic, how it was implemented, what assumptions about the implementation bothered, etc.
However, our participants \textit{implied} two ways of mapping (that is, associating different entities of interest with relations) that could bring some structure.


\smallskip
\noindent \textbf{Mapping Imperfect Conditions to Values.}
The first mapping our participants implicitly performed or suggested was to establish how imperfect conditions of data or modeling conditions affected the realization of abstract values.
Several participants (n=6) found that such a mapping would help them choose the feasible alternative among different realizations.
For instance, P02, for a collaborative project with the government on a resource allocation problem, shared that several conditions impacted their operationalization of the ``equitable'' allocation of resources. 
After they received a vague high-level requirement to build models for the equitable distribution of a state resource, P02 had to infer what equity meant according to their client's project requirements and analyzed several realizations of equity: resources reaching recipients on time, geography, representation of different receiving communities, etc.
Each of these realizations was impacted by several imperfect conditions, such as missing information on community representations and inaccurately collected geographical markers.
P02 and their team had unstructured discussions to create a map of how different conditions impacted their realizations of values to choose a path where addressing the imperfections was feasible. 
Processes like the above were followed informally by many practitioners (n=9). 

Further, our participants also found that mapping imperfections to values could help keep track of issues arising from imprecise value operationalizations.
As discussed in section \ref{sec:constraints} on how top-down institutional factors constrained methodological choices, some participants shared that they had to compromise on accurately using ethical toolkits to align with business values.
For instance, P11 shared that they broke several assumptions of SHAP \rev{(an XAI tool, \cite{lundberg2017unified})} to satisfy the transparency value of their organization:

\begin{quote}
    \textit{``So we have to sometimes break some of the assumptions of SHAP; for instance, they would expect us to run SHAP with, say, 40 or 50 data points. And yeah, so when the assumptions don't hold, then the features that SHAP deemed to be important also have to be taken with a good grain of salt.''}
\end{quote}



\smallskip
\noindent \textbf{Mapping Interventions to Imperfect Conditions.}
Similar to the previous mapping, the association of interventions used to address various imperfect conditions has many-to-many characteristics.
Though mostly technical, participants typically employ different strategies to address an imperfection, but not all paths were recorded correctly or communicated to different stakeholders.
For instance, P15 shared that they took several steps--such as removing certain features, randomizing features, and collecting new information---to address bias against a particular group. However, some details on randomization were not discussed with domain experts, assuming that the interpretation of randomized features and their association with model outcomes would not change.
In the case of P09, third-party data was used to address both imbalanced data issues and distribution shifts.
However, since data shift was also addressed by designing better learning algorithms, the contribution of how new data improved data shift issues was not documented.
Practitioners, like P09, shared that the model cards they used were very descriptive and highlighted a need to show how their interventions map to addressing different imperfect conditions. 


Our participants also found that the above mapping can improve their accountability practices to end users and other external stakeholders. 
P02 shared that their small team of five developers made several decisions on defining geographical boundaries when annotating satellite images for a project on deciding which geographical locations needed the most roads.
However, they also noted that including such decisions they took to address various imperfections and the shortcomings of each of these paths could help them communicate their decisions to different stakeholders and be more accountable:

\begin{quote}
    \textit{``Ideally, we would want polygons on the satellite images instead of a single point [to decide the areas that needed most roads]. So essentially, the training data was entirely built upon our assumptions of what a right polygon is, and the danger with that is that people at the margins stand a chance of being excluded because the annotation was being done by four or five people. If our polygon excluded a piece of land, but the people there needed roads, we had no clue to know that information to change our polygon shapes.''}
\end{quote}




\section{Discussion}
\rev{Our findings show that technical practitioners' methodological approach to RML falls along a spectrum of idealization.
It is important to remember that practitioners do not deliberately follow or implement ideal or non-ideal theories.
Instead, if we take an ideal-non-ideal methodological lens, some of their approaches align with ideal modes of theorizing justice or fairness (section \ref{sec:ideal}), whereas others sympathize with non-ideal theory (section \ref{sec:nonideal}).
Altogether, though the fields of HCI and ML have produced several toolkits, frameworks, and guidelines that technical practitioners use at various stages of an ML lifecycle, mostly in isolation (section \ref{sec:rel_2.1}),
our findings highlight that they need overall methodological support to systematically map their assumptions and choices (to perform an intervention) with the intended values of their system.}

\rev{Below, we discuss a new methodological framework to facilitate collaboration between technical practitioners (whose subtle choices and assumptions are under-scrutinized in prior HCI literature) and other stakeholders of an ML lifecycle. We see two use cases of adopting this framework:}
\begin{enumerate}
    \item \rev{Besides invoking critical reflection to identify gaps and discuss potential interventions, our framework also provides a language to communicate under what realities the interventions work and how they map to intended values. It further facilitates technical practitioners' collaboration with other stakeholders who contribute to value framing (such as managers) and shed light on assumptions under which interventions are enacted (such as domain experts).}
    \item \rev{Our framework can serve as a public-facing accountability tool, such as a model card, to demonstrate various methodological choices and assumptions made to realize the values of the overall system.}
\end{enumerate}
\rev{We develop this framework by referring to concepts from non-ideal theorizing of justice to approach ML by taking a stance sufficiently sensitive to real-world complexities that practitioners face.
Section \ref{sec:map1} describes the components that constitute our framework, and section \ref{sec:map2} discuss how to approach RML using our framework.
}



\subsection{Components of Mapping}
\label{sec:map1}
Our framework first suggests practitioners identify five types of components and then establish relations between them. 
Components here refer to states of modeling (undesirable data properties and realized goals), actions taken during modeling (interventions), conceptions (abstract values), and known assumptions. We describe each of them in order.

\smallskip
\noindent \textbf{Undesirable Properties (UP).}
We use this term to refer to any indivisible properties of the data that practitioners believe will hinder achieving their objectives of using ML.
Sometimes, a clear articulation of modeling objectives is required to identify UP and their impact on objectives.
At other times, a less sophisticated and incomplete picture of objectives is sufficient to diagnose UP.
We discuss how to articulate the objectives of ML systems using Abstract Values (AV) and Realized Goals (RG) in the section below.
Note that our definition of UP is data-centric \cite{jarrahi2022principles}: though several decisions in model training and evaluation affect model objectives, any modeling decisions and steps taken are actions performed by practitioners, which we discuss under Interventions (IV) below.

We will now unpack our definition of UP with a few examples. While some data properties that affect modeling are immutable throughout the ML lifecycle, others are transformable.
Consider a dataset with a large number of categorical and ordinal features with high cardinality. This high cardinality is immutable and is also an undesirable data property because it is challenging to encode such features to a real-valued vector space without losing some information, which could have fairness implications (for example, see \cite{kulkarni2022predicting}).  
Most data properties are transformable and typically are unreflective of real-world context. For instance, construct-feature mismatch, missing data, and unbalanced labels describe undesirable properties that are related to fairness, and several techniques are followed in practice to approach these conditions \cite{fernando2021missing,subramanian2021fairness,caoHeteroskedasticImbalancedDeep2021}.
Note that our definition of an undesirable data property is indivisible and specific so that interventions attempt to address them clearly. For instance, data drift or concept drift cannot be specified as undesirable since they can occur for various reasons. Instead, specific causes of these issues, such as unrepresentative data concerning a population subset, should be explicitly specified as an undesirable property.

\smallskip
\noindent \textbf{Abstract Values (AV) and Realized Goals (RG).}
AV are high-level abstract values or ideals practitioners want their ML systems to adhere to. Some examples include accuracy, fairness, transparency, robustness, and inclusivity. The boundaries between AV are typically nebulous. Realized Goals are what practitioners will achieve in relation to data and/or models and \textit{hypothesize} to be realizations of their abstract values. 
We note that RG are hypothesized realizations of AV since there are no ways to prove these realizations are true objectively.
Examples of RG can be as simplistic as satisfying a fairness metric to realize fairness or can involve realizing temporal robustness through a complex design of experiments and evaluating model performance on different cohorts.
Together, AV and RG describe what practitioners intend to achieve with ML: while AV express their objectives abstractly, RG are what they actually achieve.

\smallskip
\noindent \textbf{Interventions (IV) and Known Assumptions (KA).}
IV are concrete actions practitioners perform to address different UP, thereby achieving RG and adhering to AV. 
Depending on how UP are diagnosed to be caused, a combination of computational or non-computing actions can be intervened. 
While examples of computational interventions include the application of fairness metrics, feature transformations, imputing strategies, and causal inference algorithms to break feedback loops, some non-computing approaches are data relabelling, improving the diversity of annotated data, and including more people in problem formulation.
While all components discussed so far are required in our framework, KA are optional and included based on contextual factors.
For instance, prior work has argued that many commonly used fairness metrics degrade the performance of a few groups to improve that of targeted groups, a phenomenon called leveling down \cite{mittelstadt2023unfairness}. Practitioners may assume that a small degradation is acceptable in their application. This is recorded under KA in our framework. 

\subsection{Procedure of Mapping}
\label{sec:map2}
Once practitioners identify the five types of components, our framework then requires them to establish relations and map these components in four phases. 

\smallskip
\noindent \textbf{Phase 1: Diagnosis of Undesirable Properties.}
Our framework begins with diagnosing undesirable data properties specific to a problem. This resonates with how non-ideal theory starts with a detailed empirical investigation of systematic injustices in the actual world and then seeks a causal chain of actions that can be enacted. 
Our framework first requires practitioners to conduct a detailed investigation of the undesirable data properties of the ML system that affect the intended objectives of building fair ML.
Some readers may argue that several modeling decisions have fairness implications and ask why those are not initially diagnosed. First, we imagine UP as describing the \textit{state} of the world where unfairness begins, similar to how injustices exist in a non-ideal world. We then imagine any modeling steps or decisions as the interventions enacted to address unfairness. We acknowledge that these interventions can make unfeasible or inaccurate assumptions and cause discrimination. However, this discrimination arises from the actions performed by practitioners (by making some assumptions), and our framework exposes them through the explicit mapping between KA, IV, and UP.

Practitioners already use a range of techniques to identify undesirable data properties. 
Our objective is not to discuss how to conduct diagnosis but to emphasize practitioners begin their RML approaches with this phase. 
For instance, practitioners use interrogative toolkits such as domain-focused checklists \cite{madaioCoDesigningChecklistsUnderstand2020,deon} or Datasheets to identify potential data biases \cite{gebruDatasheetsDatasets2021,papakyriakopoulos2023augmented}, use rule-based functional tests to debug and find errors caused by data \cite{ribeiro2020beyond,rottger2020hatecheck}, conduct counterfactual experiments to detect discriminatory properties \cite{wu2021polyjuice,kommiya2021towards}, and even use visualization-based tools to inspect data bias visually \cite{robertson2023angler,denkowski2012challenges}.
While some data issues are apparent, sometimes, practitioners uncover undesirable properties during model training and assessment. For instance, several participants in our study shared that they use explainability toolkits to conduct error analysis and surface data biases.
However, our findings showed that practitioners already pursuing this phase do not follow a structured and accountable process. This methodological framework requires diagnosing UP systematically \textit{in relation} to fairness and ethics.

\smallskip
\noindent \textbf{Phase 2: Mapping Undesirable Properties, Realized Goals, and Abstract Values.}
In this phase, our framework requires practitioners to carefully map how each diagnosed undesirable data property affects the realization of tangible goals, thereby impacting adhering to abstract values/ideals.
Either RG or AV can be constructed first in practice. Several constraints we discussed in our findings, such as top-down institutional structure or ideological factors, can determine if AV or RG is construed first. Regardless, the direction of mapping is always from RG to AV, and this relationship is many-to-many in our framework: achieving one or more realizable goals can imply hypothetically adhering to an abstract value, or achieving one realizable goal can be a factor in hypothetical adherence of an abstract values. 
Similarly, a mapping from UP to RG is also many-to-many, implying that several undesirable data properties affect realizing several tangible goals.
We discussed that diagnosing UP might sometimes require a clear articulation of the intended objectives of an ML system.
Our framework provides a structured way of thinking about these model objectives: once a rough mapping between RG and AV is enacted, diagnosing UP in relation to intended objectives becomes straightforward and structured.
Practitioners still primarily diagnose UP initially, where the AV-RG mapping simply supports the process.

\citet{wiens2015against} is critical of ideals or abstract values as useless for deciding political actions in a non-ideal world.
However, other non-ideal theorists, such as \citet{anderson2010imperative}, describe ideals as imagined solutions to identified societal problems, where they are hypotheses that can be tested, evolved, and re-conceived.
We align with \citet{anderson2010imperative}'s take on non-ideal theory and consider the realizations of AV into RG as hypotheses.
\rev{This is because} we observe that the operationalization between AV and RG exists in the mental models of practitioners, and hence, RG are what practitioners hypothesize to be realizations of their abstract values or ideals.
Further, these hypothesized realizations are challenged by UP, and some notion of AV-RG map helps practitioners diagnose and address these UP.
This also aligns with the realist approach to injustice \cite{cozzaglio2022feasibility,favara2023political}, where some reference to ideal desirability is required to justify feasible political actions.
In the case of ML, we discussed that having a rough sense of the AV-RG map helps practitioners to surface a few UP that are otherwise difficult to identify and address. 

For example, consider transparency a value that a practitioner hypothesizes to adhere to by developing local and global explanation models. These models are our practitioner's hypothesized realizations of transparency and do not imply transparency is achieved; they are hypotheses that need to be evaluated and re-conceived if they do not solve real problems.
Further, our practitioner diagnoses various UP of data, such as feature dependencies and class imbalance, which affect their RG and AV. If our practitioner had imagined different RG for transparency other than model explanations, perhaps feature dependencies might not have been taken into account. 
Overall, this mapping between UP, RG, and AV provides practitioners and other stakeholders a blueprint of how different UP affect their hypothesized realization of ideals.


\smallskip
\noindent \textbf{Phase 3: Mapping Known Assumptions, Interventions, and Undesirable Properties.}
In this phase, practitioners map their interventions to different UP to denote how one or more interventions handled an indivisible and specific data property. Some interventions work under assumptions, and our framework requires practitioners to make these associations explicit. Note that in ideal theory, interventions are directly associated with abstract values. For instance, satisfying fairness metrics is equated to fairness. However, in our framework, the extent to which an intervention addresses an undesirable property depends on the assumptions under which the intervention is enacted. We also recognize that these interventions and assumptions are a function of various constraints we discussed in our findings. 
All these characteristics align with a realist and non-ideal perspective of pursuing fairness as a fact-sensitive vocation \cite{wiens2015against,cozzaglio2022feasibility}.

\smallskip
\noindent \textbf{Phase 4: Iteration and Versioning.}
Our framework is a work in non-ideal theory sensitive to real-world facts and constraints. We recognize that the mapping between the abovementioned components will not be completed in a single session but will evolve. Some mapping may change, and some values may become irrelevant or inappropriate. Hence, our framework requires practitioners to iterate the above three phases until a desirable state is reached or until saturation. 

Further, in our interviews, several participants highlighted that versioning various fairness-related decisions is required to understand how a particular path was reached.
So, in this framework, we need our practitioners to maintain a record of deleted components and associations, and for each component, practitioners should be aware of how the mapping from that component has changed over time.
If we imagine our framework as a network of nodes and connections, the versioning is how a node has changed its outward connections over time. We leave such explorations on how versioning can be presented to future work.


\subsection{Practical Considerations and Limitations}
\label{sec:map3}
\begin{figure}[t]
    \centering
    \includegraphics[scale=0.45]{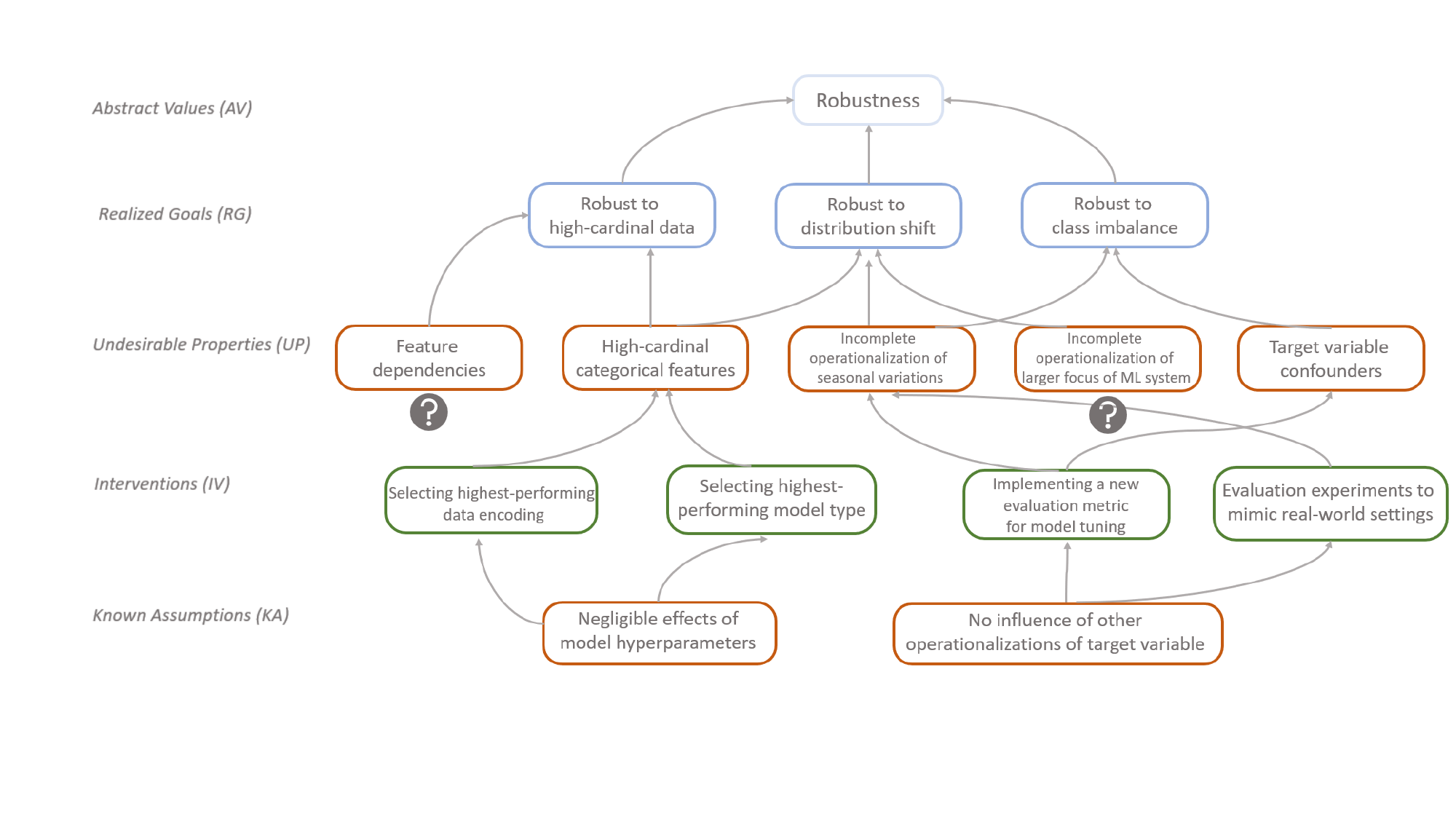}
    \caption{An imagination of how our methodological framework can be presented. To illustrate, consider 'Feature dependencies' as an undesirable property mapped to the realized goal, 'Robust to high-cardinal (categorical) data.' This implies feature dependencies affect achieving the goal of building robust models with high-cardinal categorical features.
    Note that such visualizations can help identify and keep track of unaddressed undesirable properties (such as ``feature dependencies'' and ``incomplete operationalization of larger focus of ML systems'' in this example) and incomplete mappings between different components (for instance, ``feature dependencies'' can affect addressing distribution shifts but are not mapped explicitly in this example).}
    \label{fig:rai-map}
\end{figure}

Our framework requires an extensive mapping between five kinds of components: undesirable data properties, interventions, known assumptions, realized goals, and abstract values.
It should be noted that we need more than just enumerating different components and establishing relations between them to complete our framework. As we emphasized in phase 4 of the mapping procedures, practitioners must iterate this process and version any changes.
We discussed that such a process would require practitioners to conduct a structured, nuanced, and deeper empirical investigation of real-world conditions.
This contrasts with the current dominant RML approach involving idealizations, abstractions, simplifications, undocumented assumptions, and unstructured procedures.

However, a critical piece not discussed in this paper is how practitioners would \textit{present} this methodological framework to other stakeholders.
Unlike the typically followed reductionist approach to fairness, our methodology requires practitioners to clearly articulate and record all assumptions, decisions, and relations between various components.
Even if an ML practitioner diligently follows our framework, clearly communicating these methodological choices to stakeholders in a consumable format is equally important. 
RML is a multi-stakeholder vocation, and discussing the methodological choices of our practitioners with different stakeholders is crucial in various aspects, including creating new components and relations between components.

Documenting all the steps is one way to go. However, such a document could soon become cumbersome to follow and get feedback.
Another approach (as we hinted towards the end of the previous section) is to visualize our framework on a web application as a dynamic network of nodes and connections. Nodes here are components, and connections refer to mappings that practitioners create.
We imagine such an application in Figure \ref{fig:rai-map} for common examples of the components we discussed.
\rev{The framework starts with UP as the central element and can be imagined to have an input and an output chain. The input chain denotes that KA affect IV which in turn affect UP (hence the flow KA→IV→UP). In the output chain, UP affect RG which in turn affect AV (hence the flow UP→RG→AV).}
While this approach has the advantage of being interactive, it needs to be clarified how to gracefully scale such a web application to a large ML system with several undesirable properties, interventions, and abstract values.
This is beyond the scope of our current work, and further research is required to investigate the design side of such a presentable framework.

\rev{We also acknowledge that our framework does not definitively account for hidden work, such as that of data annotators, that goes into building ``responsible'' ML systems. In some cases, when annotation work is done through partnerships with specific communities (e.g., \cite{mokhberi2023development}), practitioners could use our framework to account for their voices. But prior research has discussed how most data annotation work creates a toxic and exploitative environment \cite{sarkar2023enough,kapania2023hunt,wang2022whose}, and we leave the creative exploration of our framework to facilitate \textit{responsible} collaboration between practitioners and exploited stakeholders, such as data labelers, to future work.}



\bibliographystyle{ACM-Reference-Format}
\bibliography{references}

\end{document}
\endinput